\documentclass[aps,prb,twocolumn,superscriptaddress]{revtex4}

\usepackage{graphicx}
\usepackage{amsmath}

\begin{document}

\title{Nonequilibrium characteristics in all-superconducting tunnel structures}

\author{M. A. Laakso}
\author{P. Virtanen}
\affiliation{Low Temperature Laboratory, Helsinki University of
Technology, P.O. Box 2200 FIN-02015 TKK, Finland}
\author{F. Giazotto}
\affiliation{NEST-CNR INFM \& Scuola Normale Superiore, I-56126 Pisa, Italy}
\author{T.~T. Heikkil\"a}
\email[]{Tero.T.Heikkila@tkk.fi}
\affiliation{Low Temperature
Laboratory, Helsinki University of Technology, P.O. Box 2200
FIN-02015 TKK, Finland}

\date{\today}

\begin{abstract}
We study the nonequilibrium characteristics of superconducting
tunnel structures in the case when one of the superconductors is a
small island confined between large superconductors. The state of
this island can be probed for example via the supercurrent flowing
through it. We study both the far-from-equilibrium limit when the
rate of injection for the electrons into the island exceeds the
energy relaxation inside it, and the quasiequilibrium limit when the
electrons equilibrate between themselves. We also address the
crossover between these limits employing the collision integral
derived for the superconducting case. The clearest signatures of the
nonequilibrium limit are the anomalous heating effects seen as a
supercurrent suppression at low voltages, and the hysteresis at
voltages close to the gap edge $2\Delta/e$, resulting from the
peculiar form of the nonequilibrium distribution function.
\end{abstract}


\maketitle

\section{Introduction}
New device concepts based on nonequilibrium effects in
superconducting mesoscopic tunnel structures have been proposed in
the last few years. These include Josephson transistors, electron
refrigerators and
thermometers.\cite{giazotto+heikkila_rev,pekola_prl} In Josephson
transistors the supercurrent flowing through a superconductor-normal
metal-superconductor (SNS) weak link can be suppressed or even
reversed in a
$\pi$-transition\cite{baselmans_nature,shaikhaidarov,huang} by
driving the normal metal part out of equilibrium through injection
of charge carriers from additional terminals. When the additional
terminals are superconductors connected by tunnel junctions, the
supercurrent can also be enhanced.\cite{giazotto+heikkila} This
transistorlike operation with large current and power gain has also
been experimentally demonstrated.\cite{savin+pekola} Also an
all-superconducting SISIS transistor in the quasiequilibrium regime
has been theoretically addressed.\cite{giazotto+pekola} In the
quasiequilibrium limit the electron-phonon interaction is nearly
absent and the sample can be considered as detached from the phonon
bath. The high frequency of electron-electron collisions still
serves as a method of relaxation and the electrons assume a Fermi
distribution but with a temperature that in general differs from the
temperature of the phonon bath. Here we study a similar SISIS
structure with arbitrary strength of the inelastic scattering
seeking ways to characterize the degree of nonequilibrium on the
system. The paper is organized as follows: The model of the SISIS
structure is presented in Sec.~2. All the relevant equations and
calculated results are presented in Secs.~3 and 4, respectively. We
finish with a summary and a discussion in Sec.~5, where we also
address briefly the feasibility of this structure as a transistor.

\section{Model}
\begin{figure}[tb]
 \includegraphics{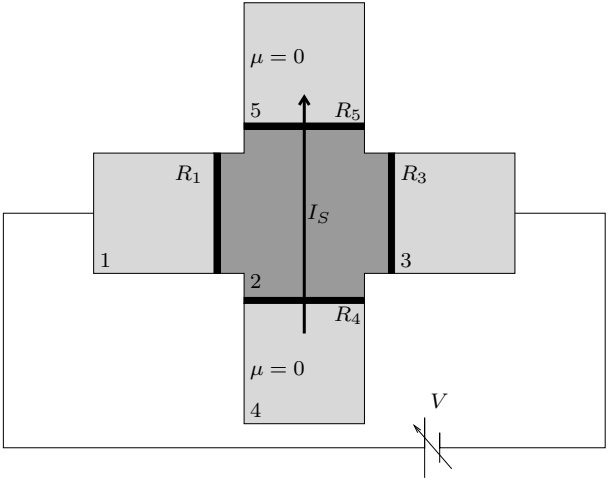}
 \caption{Scheme of the SISIS structure studied in this work. The superconducting island (2) in the middle is connected with tunnel contacts to four large superconducting leads (1,3,4 and 5). The control line is biased with voltage $V$, which controls the energy distribution on the island. A supercurrent $I_S$ is driven across the island from lead 4 to lead 5, and its magnitude depends on the distribution function on the island. Each SIS junction is a tunnel contact of resistance $R_i$.}
 \label{fig:structure}
\end{figure}

The superconducting structure under study is schematically depicted
in Fig.~\ref{fig:structure}. We characterize the mean free path that
the electron travels before scattering by scattering lengths
$l_{el}$ for elastic scattering and $l_{e-ph}$ and $l_{e-e}$ for
inelastic electron-phonon and electron-electron scattering,
respectively. In mesoscopic systems typical orders of magnitude are
$l_{el}\approx10\ldots100\:\mathrm{nm}$ and
$l_{e-e}\approx1\ldots20\:\mathrm{\mu m}$. The electron-phonon
scattering length depends strongly on temperature. For a typical
copper wire $l_{e-ph}\approx21\:\mathrm{\mu m}$ at 1 K but at 100 mK
we already have $l_{e-ph}\approx670\:\mathrm{\mu
m}$.\cite{giazotto+heikkila_rev} In other metals these length scales
are of the same order of magnitude. The superconducting island in
the middle is assumed to have small dimensions so that $L \ll
l_{e-e},l_{e-ph}$, leading to weak energy relaxation via inelastic
scattering. As we shall see in the following, in this case it is
possible to drive the electron energy distribution of the island out
of equilibrium by quasiparticle injection from the superconducting
leads. The degree of nonequilibrium on the island can then be probed
for instance by measuring the supercurrent driven through the island
via an additional SISIS line. The leads are assumed to remain in
thermal equilibrium due to their large dimensions. We further assume
that the resistances of the tunnel contacts are large compared to
the normal state resistance of the superconducting island. This
allows us to use the tunnel Hamiltonian approach, in which each
region has spatially constant, separate energy distributions,
independent of the directions of the momenta.

\section{Formalism}
\subsection{Green's functions in SISIS structure}
We use the quasiclassical Keldysh Green function formalism together with the tunnel Hamiltonian model in describing our system. It has previously been successfully applied to hybrid structures with normal metal and superconducting islands.\cite{brinkman_etal,voutilainen} The quasiclassical Green functions in Nambu space can be written in a matrix form as
\begin{equation}\label{eq:green_function} \hat{g}=\left(
 \begin{array}{@{}cc@{}}
  g & f \\
  -f^\dagger & \bar{g}
 \end{array} \right),
\end{equation}
where $\hat{g}$ is either retarded (advanced), $\hat{g}^{R(A)}$, or
Keldysh, $\hat{g}^K$, Green's function. In the tunnel Hamiltonian
model Green's functions are isotropic with respect to directions of
the momenta. In this case the retarded (advanced) functions satisfy
the steady-state Usadel equations\cite{voutilainen,usadel} with
solutions
\begin{align}
 g^{R(A)}=-\bar{g}^{R(A)}=&\pm\dfrac{E\pm i\gamma}{\sqrt{(E\pm i\gamma)^2-(|\Delta|\pm i\delta)^2}}\nonumber \\ \label{eq:greenfunction}
 f^{R(A)}=f^{\dagger R(A)}=&\pm\dfrac{|\Delta|\pm i\delta}{\sqrt{(E\pm i\gamma)^2-(|\Delta|\pm i\delta)^2}}.
\end{align}
Here $\Delta$ is the superconducting order parameter, $\gamma=\sum_j
\eta_jg^{R(A)}_j$ and $\delta=\sum_j \eta_jf^{R(A)}_j$. The index
$j$ runs over the other parts of the structure that are connected
with tunnel contacts to the region in question. The characteristic
tunneling rate $\eta$ between superconductors is defined as
$\eta_j=(4\nu e^2\Omega R_j)^{-1}$, where $\nu$ is the normal-state
density of states at the Fermi level and $\Omega$ is the volume of
the island. In the tunneling limit $\eta\ll\Delta$ and we may
neglect the exact forms of $\gamma$ and $\delta$ and instead use
some constant $\gamma$ and $\delta$ in the numerical simulations.
Below, we choose $\gamma=10^{-4}$ and $\delta=10^{-5}$. This value
of $\gamma$ has been experimentally verified in Ref.~2. The Keldysh
Green function for the system can be written with the standard
parametrization as
\begin{equation}
\hat{g}^K=\hat{g}^R(f_L+\hat{\tau}_3f_T)-(f_L+\hat{\tau}_3f_T)\hat{g}^A,
\end{equation}
where $\hat{\tau}_3$ is the third Pauli spin matrix. Here we have
also used the odd- and even-in-$E$ parts of the distribution
function:
\begin{align*}
 f_L(E)=&-f(E)+f(-E), \\ f_T(E)=&1-f(-E)-f(E).
\end{align*}
The full distribution function can be recovered with $2f(E)=1-f_L(E)-f_T(E)$. We also define
\begin{align}
 g^{(-)}&=\mathrm{Re}\:g^R=\dfrac{1}{2}(g^R-g^A), \nonumber \\
 f^{(-)}&=\mathrm{Re}\:f^R=\dfrac{1}{2}(f^R-f^A), \nonumber \\ f^{(+)}&=\mathrm{Im}\:f^R=\dfrac{1}{2i}(f^R+f^A). \nonumber
\end{align}
The functions $f^{(+)}$ and $g^{(-)}$ are even in $E$ and $f^{(-)}$ is odd. The density of states is given by $g^{(-)}$. The odd and even parts of the nonequilibrium distribution function can now be found from the kinetic equations presented in Ref.~9. The resulting equations are
\begin{widetext}
\begin{align}
 -4J_1\nu_2e^2\Omega_2=&g^{(-)}_{2,E}G_1\left\lbrace g^{(-)}_{1,E+\mu_1}\left(f_{L2}+f_{T2}-f_{L1}-f_{T1}\right)+g^{(-)}_{1,E-\mu_1}\left(f_{L2}-f_{T2}-f_{L1}+f_{T1}\right)\right\rbrace \nonumber \\ &+g^{(-)}_{2,E}G_3\left\lbrace \label{eq:kineticeqL} g^{(-)}_{3,E+\mu_3}\left(f_{L2}+f_{T2}-f_{L3}-f_{T3}\right)+g^{(-)}_{3,E-\mu_3}\left(f_{L2}-f_{T2}-f_{L3}+f_{T3}\right)\right\rbrace, \\
 \left(8|\Delta_2|f_{T2}f^{(+)}_2-4J_2\right)\nu_2e^2\Omega_2=&g^{(-)}_{2,E}G_1\left\lbrace g^{(-)}_{1,E+\mu_1}\left(f_{L2}+f_{T2}-f_{L1}-f_{T1}\right)+g^{(-)}_{1,E-\mu_1}\left(-f_{L2}+f_{T2}+f_{L1}-f_{T1}\right)\right\rbrace \nonumber \\ &+g^{(-)}_{2,E}G_3\left\lbrace g^{(-)}_{3,E+\mu_3}\left(f_{L2}+f_{T2}-f_{L3}-f_{T3}\right)+g^{(-)}_{3,E-\mu_3}\left(-f_{L2}+f_{T2}+f_{L3}-f_{T3}\right)\right\rbrace, \label{eq:kineticeqT}
\end{align}
\end{widetext}
where $G_i=1/R_i$ are the conductances of the tunnel contacts, $\mu_i$ are the chemical potentials of the regions $i$ and $J_i$ are the collision integrals for the energy relaxation.

\subsection{Order parameter and currents}
The pair potential in the central island must be solved self-consistently from the equation
\begin{equation}\label{eq:orderparameter}
 |\Delta_2|=\frac{\lambda}{2}\int^{E_C}_{-E_C}dE f_{L2}f^{(-)}_E,
\end{equation}
where $E_C$ is the BCS cutoff energy and $\lambda$ is the electron-electron interaction parameter. When a SIS-junction is not biased with an external voltage, the supercurrent flowing across the junction is given by
\begin{align}
 I^{2\to4}_S=&-\dfrac{1}{2eR_4}\displaystyle\int^\infty_{-\infty} dE \left\lbrace \left(f_{L2} f^{(-)}_2\:f^{(+)}_4+f_{L4} f^{(-)}_4\:f^{(+)}_2\right)\right.\nonumber \\ \times&\sin(\chi_4-\chi_2)+(f_{T2}-f_{T4})\left(g^{(-)}_2g^{(-)}_4+f^{(+)}_2\:f^{(+)}_4\right.\nonumber \\  &\times\left.\Bigl.\cos(\chi_4-\chi_2)\Bigr)\right\rbrace.\label{eq:supercurrent}
\end{align}
The first part of the equation multiplying the sine term represents
the usual dc-Josephson relation where $\chi_{2,4}$ are the
macroscopic phases of the respective superconductors. The term
$f^{(-)}_2\:f^{(+)}_4$ is finite only between $\Delta_2<E<\Delta_4$
whereas the term $f^{(-)}_4\:f^{(+)}_2$ is finite when
$\Delta_4<E<\Delta_2$. The second part in
Eq.~\eqref{eq:supercurrent} usually vanishes because $f_T=0$ in
quasiequilibrium. If a finite charge imbalance develops on the
island, $f_{T2}$ deviates from zero, and the second part contributes
to the current as well.

If a voltage is applied across the junction the phase difference begins to evolve in time and the supercurrent averages to zero. In this case only the tunneling of quasiparticles contributes to the current, so that we have
\begin{align}
 I^{1\to2}=&-\dfrac{1}{4eR_1}\displaystyle\int^\infty_{-\infty} dE \left\lbrace g^{(-)}_{1,E+\mu}g^{(-)}_{2,E}\left(f_{L2}+f_{T2}\right.\right.\nonumber \\ &\left.-f_{L1}-f_{T1}\right)+g^{(-)}_{1,E-\mu}g^{(-)}_{2,E}\left(-f_{L2}+f_{T2}\right.\nonumber \\ &\Bigl.\left.+f_{L1}-f_{T1}\right)\Bigr\rbrace. \label{eq:qpcurrent}
\end{align}
The quasiparticles tunneling through the junction also carry heat. The energy current is
\begin{align}
 I^{1\to2}_E=&-\dfrac{1}{4e^2R_1}\displaystyle\int^\infty_{-\infty} dE E\left\lbrace g^{(-)}_{1,E+\mu}g^{(-)}_{2,E}\left(f_{L2}+f_{T2}\right.\right.\nonumber \\ &\left.-f_{L1}-f_{T1}\right)+g^{(-)}_{1,E-\mu}g^{(-)}_{2,E}\left(f_{L2}-f_{T2}\right.\nonumber \\ &\Bigl.\left.-f_{L1}+f_{T1}\right)\Bigr\rbrace, \label{eq:energycurrent}
\end{align}
which is used in determining the electron temperature in quasiequilibrium.

\subsection{Energy relaxation}
In practice the inelastic scattering is never completely absent. At
low temperatures the most relevant relaxation mechanism is
electron-electron scattering, which can be included with $e-e$
collision integrals. We may also study cases where the dimensions of
the island are no longer significantly smaller than the
electron-electron scattering length, i.e., $L \lesssim l_{e-e} \ll
l_{e-ph}$. The collision integral for a screened Coulomb interaction
in a diffusive wire is known\cite{altshuler_aronov} and has been
used in the analysis of a SINIS system.\cite{giazotto+heikkila} It
is strictly valid only for a normal metal island, however. To get a
qualitative picture of the changes due to superconductivity in
energy relaxation, we apply instead a collision integral where the
structure of the Nambu space has been taken into account. In the
clean limit the potential of a distant electron is completely
screened by all other electrons in the superconductor and the
electron-electron interaction can be approximated by a point
interaction. In this case the potential may be modelled with a delta
function $V(\mathbf{r})=\nu_2 \lambda_{ee}\delta(\mathbf{r})$ and
the collision integral is \cite{kopnin_tons}
\begin{align}
 J^{(ee)}_1(E_3)=&\kappa\iint dE_1 dE_2 \left\lbrace \left(g^{(-)}_{E_1}g^{(-)}_{E_2}-f^{(-)}_{E_1}f^{(-)}_{E_2}\right)\right.\nonumber \\ \times&\left(g^{(-)}_{E}g^{(-)}_{E_3}+f^{(-)}_{E}f^{(-)}_{E_3}\right)\Bigl(\left(1-f_{E}\right)f_{E_1}f_{E_2}f_{E_3}\Bigr.\nonumber \\ &\Bigl.\left.-f_{E}\left(1-f_{E_1}\right)\left(1-f_{E_2}\right)\left(1-f_{E_3}\right)\Bigr)\right\rbrace,
\end{align}
where $\kappa=4\lambda_{ee}^2\pi/p_Fv_F$, $v_F$ and $p_F$ are the
Fermi velocity and momentum, respectively, and energies satisfy the
conservation law $E=E_1+E_2+E_3$. The second collision integral,
$J^{(ee)}_2$, vanishes in a left-right symmetric structure. We note
that because the terms $g^{(-)}_{E_3}$ and $f^{(-)}_{E_3}$ in the
kernel of the integral assume very small values when $|E_3|<\Delta$,
the collision integral has a very small effect on excitations inside
the gap. On the other hand, energy relaxation is strongest for
excitations at $|E_3|=\Delta$ due to sharp peaks at the edge of the
gap in these same terms.

\section{Results}
\subsection{Full nonequilibrium}
We begin by presenting the calculated distribution function along with the order parameter and electric currents for the simplest, namely left-right symmetric case, where the tunnel junction resistances are the same and reservoirs 1 and 3 are similar superconductors, i.e., $R_1=R_3=R$ and $|\Delta_1|=|\Delta_3|=|\Delta_L|$. When the structure is biased with a voltage $V$, the conservation of electric current forces the chemical potentials of reservoirs 1 and 3 to $\mu_1=eV/2$ and $\mu_3=-eV/2$, respectively.
\subsubsection{Distribution function}
\begin{figure}[tb]
\includegraphics[width=8.0cm]{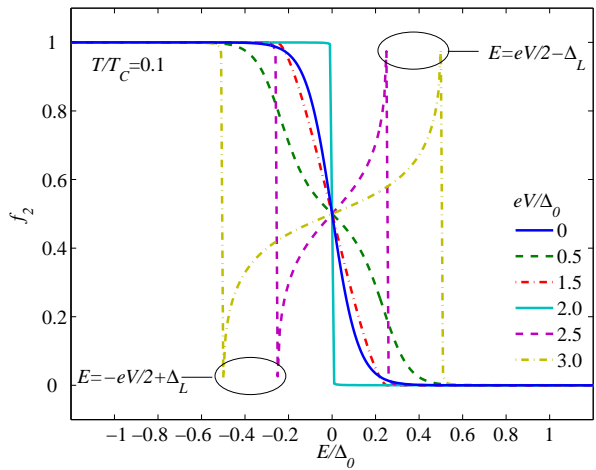}
\caption{(Color online) Nonequilibrium distribution function for the superconducting island at $T=0.1\:T_C$. The cooling effect reducing the number of excited quasiparticles as the voltage is increased is evident. Here and below we denote $\Delta_0=\Delta_L(T=0)$ and $T_C$ is the critical temperature of the leads.}
\label{fig:noneqdist}
\end{figure}

\begin{figure}[tb]
\includegraphics[width=8.0cm]{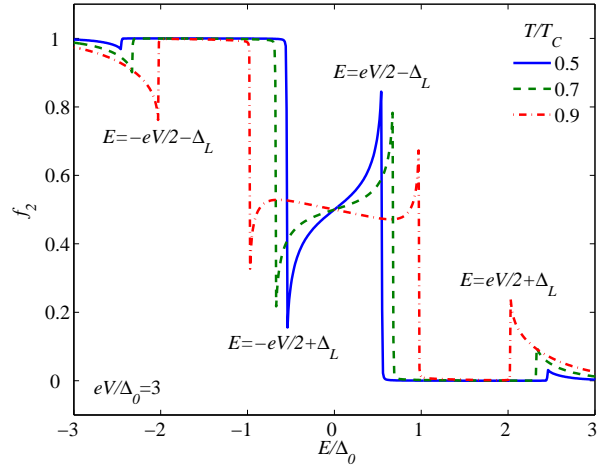}
\caption{(Color online) Nonequilibrium distribution function for the superconducting island at various bath temperatures for a voltage $eV/\Delta_0=3$.}
\label{fig:noneqdisttemps}
\end{figure}

The solution of the kinetic equations \eqref{eq:kineticeqL} and \eqref{eq:kineticeqT} in the absence of energy relaxation ($J_1=J_2=0$) may be written in terms of the full distribution functions as
\begin{equation}
 f_2=\frac{g^{(-)}_{E+\mu}f_1+g^{(-)}_{E-\mu}f_3}{g^{(-)}_{E+\mu}+g^{(-)}_{E-\mu}}.
\end{equation}
This form is remarkably simple due to the symmetry of the problem,
and it can also be derived by considering the conservation of
electric current.\cite{heslinga+klapwijk} The distribution function
is plotted in Fig.~\ref{fig:noneqdist} for various bias voltages at
a bath temperature of $0.1\:T_C$. The critical temperature of the
superconductor is $T_C=(1.764k_B)^{-1}\Delta_0$. With a small
voltage bias fewer of the states below the Fermi level are occupied
whereas the occupation is increased above the Fermi level. This
increase in excited quasiparticles can be interpreted as a heating
of the island. This anomalous heating effect stems from the
assumption of a finite $\gamma$ in Eq. \eqref{eq:greenfunction},
i.e., from the presence of quasiparticle states within the gap. In
the absence of these states, no anomalous heating is  observed. Once
the voltage is increased above $eV=\Delta_L$, the number of excited
quasiparticles on the island begins to decrease due to extraction to
states right above the energy gap in the superconducting reservoirs.
This cooling effect is discussed in Refs. 1 and 2. In
Fig.~\ref{fig:noneqdisttemps} the distribution function is plotted
at higher bath temperatures for a bias voltage $eV/\Delta_0=3$. At
these temperatures the reservoirs have more excited quasiparticles
above and below the gap, and the small notches at
$|E|=eV/2+\Delta_L$ are a result of their injection.

\subsubsection{Order parameter}
In order to measure the degree of nonequilibrium on the island we
must look for nonequilibrium induced effects in some measurable
quantities, e.g., supercurrent through the island. First we
calculate the magnitude of the order parameter with the
self-consistency equation \eqref{eq:orderparameter}. In general this
must be solved numerically.
\begin{figure}[tb]
\includegraphics[width=8.0cm]{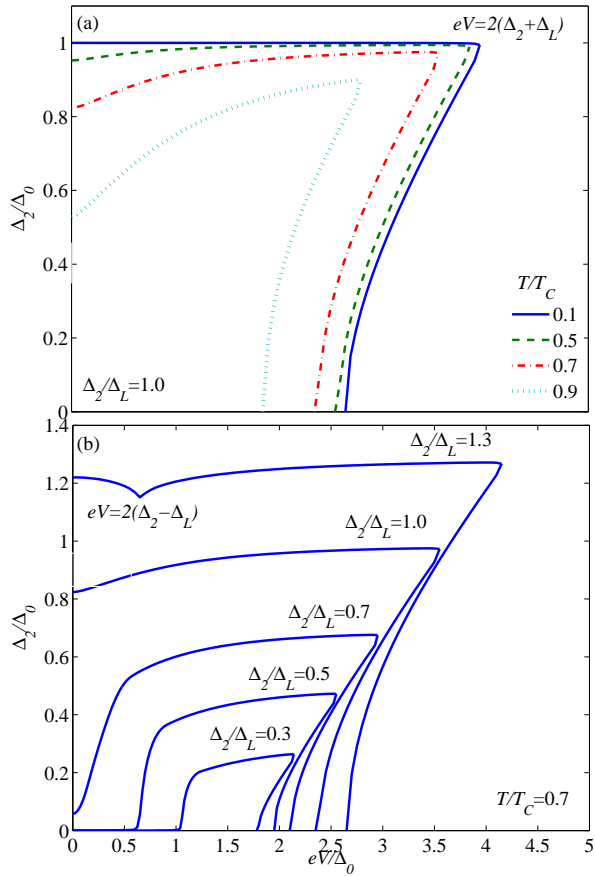}
\caption{(Color online) Order parameter as a function of bias voltage at various bath temperatures (a) and ratios $\Delta_2/\Delta_L$ (b).}
\label{fig:noneqdelta}
\end{figure}
The magnitude of the order parameter of the island as a function of
voltage at various bath temperatures is shown in
Fig.~\ref{fig:noneqdelta}(a). At $T=0.1\:T_C$ the odd-in-$E$ part of
the distribution is effectively unchanged outside the gap giving the
same result as for equilibrium. However, once $eV\gtrsim2\Delta_L$
the peculiar shape of the distribution makes it possible to have a
lower value solution for the order parameter as well, giving rise to
a hysteretic behavior with three solutions. Once voltage reaches
$eV=2(\Delta_2+\Delta_L)$ only the smallest solution, namely
$\Delta_2=0$, is possible. This is due to the fact that the order
parameter can never exceed its zero-temperature value. The
multivalued behavior of the order parameter can be interpreted as
different minima and maxima in the free
energy.\cite{giazotto+heikkila,heslinga+klapwijk} In this case the
largest and smallest values represent minima and the middle value
represents a maximum. If we increase the voltage from zero, the
system stays in the free-energy minimum corresponding to a
superconducting state. Once we enter the hysteretic region, thermal
fluctuations may cause the system to jump to normal state, which is
the other free-energy minimum. In the absence of fluctuations, the
system finally jumps to the normal state at
$eV=2(\Delta_2+\Delta_L)$. If we now proceed by decreasing the
voltage, the jump to the superconducting state may again occur
somewhere in the hysteretic region. Once the voltage is decreased
enough, only the superconducting state is possible.

At higher bath temperatures the order parameter is initially in its
equilibrium value, but increases along the voltage as the island
cools. In Fig.~\ref{fig:noneqdelta}(b) the order parameter is shown
at $T=0.7\:T_C$ but for different zero-temperature ratios
$\Delta_2/\Delta_L=T_{C_2}/T_C$. For ratios $\Delta_2/\Delta_L<0.7$,
the island is initially in the normal state because the bath
temperature is above its critical temperature. Upon increasing the
voltage, the island turns superconducting once the electron
distribution has features sharp enough to support an energy gap.

\subsubsection{Electric currents}
\begin{figure}[tb]
\includegraphics[width=8.0cm]{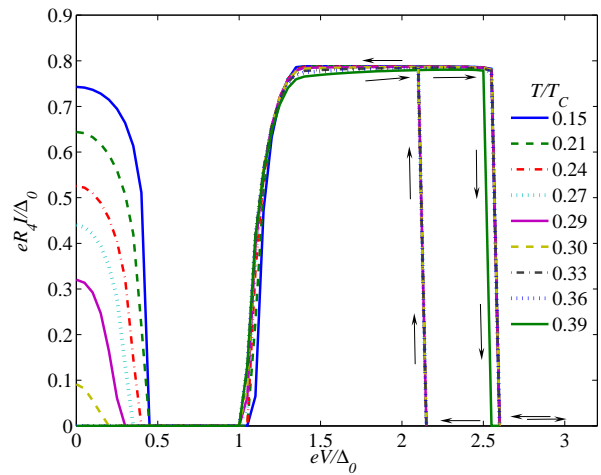}
\caption{(Color online) Supercurrent through the island in full
nonequilibrium as a function of bias voltage at various bath
temperatures with a ratio $\Delta_2/\Delta_L=0.3$. The arrows
indicate the direction the curve is traced when the bias voltage is
varied. Thermal fluctuations may cause the discontinuous jump to
occur somewhere in between the two extremes shown in the figure. The
system is assumed symmetric, i.e., $R_1=R_3=R_4=R_5$ and
$\Delta_4=\Delta_5=\Delta_L$.} \label{fig:noneqsc}
\end{figure}

\begin{figure}[tbh]
\includegraphics[width=8.0cm]{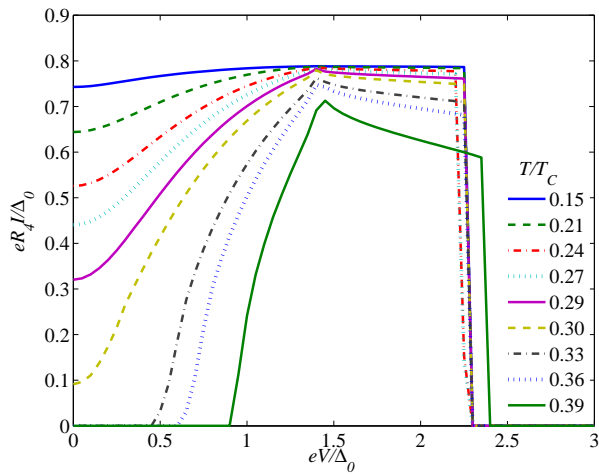}
\caption{(Color online) Supercurrent through the island in quasiequilibrium as a function of bias voltage with parameters identical to the full nonequilibrium case presented in Fig.~\ref{fig:noneqsc}.}
\label{fig:quasieqsc}
\end{figure}

Now we examine the effect that the magnitude of the order parameter has on the electric current driven through the island. In light of the results in the previous subsection the measurements should be made at relatively high temperature in order to fully bring out the variation in the energy gap. By choosing a setup with a lower $\Delta_2/\Delta_L$ ratio enables us to use a lower absolute temperature and thereby also minimize the electron-phonon relaxation, because the power injected into the phonons depends on temperature as $T^5$.\cite{wellstood} The supercurrent through the island is calculated with Eq.~\eqref{eq:supercurrent} and it is presented in Fig.~\ref{fig:noneqsc} for various temperatures assuming a ratio $\Delta_2/\Delta_L=0.3$, which corresponds roughly to the Ti/Al combination. When the bath temperature is lower than the critical temperature of the island, the initial heating effect with low bias voltages is evident. In the cooling regime the bath temperature has a negligible effect on the magnitude of the supercurrent. The hysteresis of the order parameter carries over to the supercurrent but no $\pi$-state is observed.
It is illustrative to compare these to the corresponding results in quasiequilibrium, where the high frequency of electron-electron collisions force the quasiparticles on the island to assume a Fermi distribution. The electron temperature in quasiequilibrium can be obtained by demanding that the energy current in Eq.~\eqref{eq:energycurrent} to the island vanishes (we also neglect the contribution of electron-phonon interaction to the energy current). The supercurrent in quasiequilibrium is shown in Fig.~\ref{fig:quasieqsc}. In quasiequilibrium the heating effect is absent and the island cools even with low voltages resulting in an increase of the supercurrent. Superconductivity is lost once the voltage exceeds $eV=2(\Delta_2+\Delta_L)$, just as in full nonequilibrium. The falling edge here is not hysteretic, however.
A further means to probe the degree of nonequilibrium is to voltage bias the second SISIS-line as well and measure the energy gap from the I-V curve. The quasiparticle current flowing through the probe junction in this case may be calculated with Eq.~\eqref{eq:qpcurrent}. The resulting I-V curve does not differ from its equilibrium shape, in which the current has a discontinuos jump at $eV=2(\Delta_2+\Delta_L)$.\cite{tinkham_its} The value of $\Delta_2$ and its hysteresis changes the voltage at which the jump is observed, however.

\subsection{Nonequilibrium with energy relaxation}
\begin{figure}[tb]
\includegraphics[width=8.6cm]{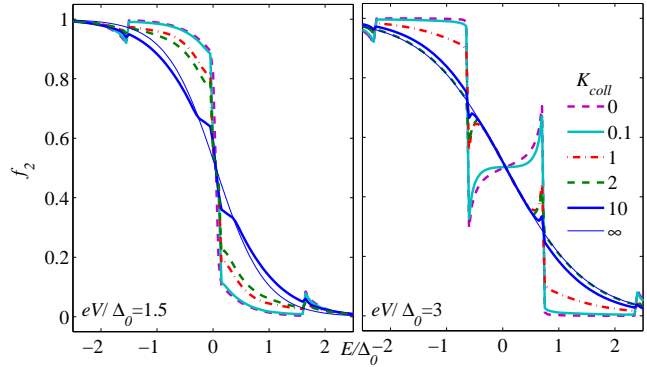}
\caption{(Color online) Distribution function for the superconducting island at $eV/\Delta_0=1.5$ (left) and $eV/\Delta_0=3$ (right) for various $\mathcal{K}_{coll}$ with $T=0.7\:T_C$. $\mathcal{K}_{coll}=\infty$ corresponds to quasiequilibrium.}
\label{fig:colldists}
\end{figure}

\begin{figure}[tbh]
\includegraphics[width=8.0cm]{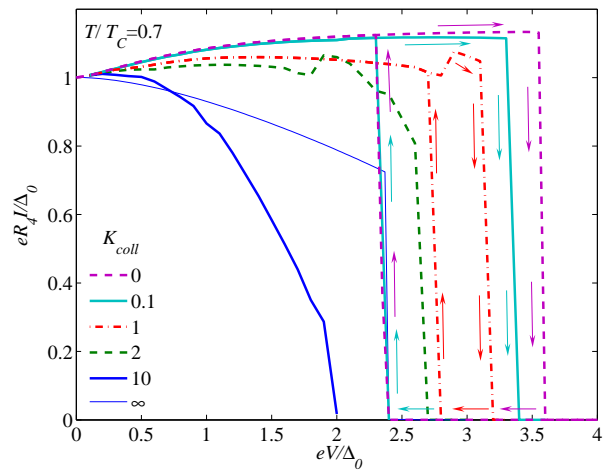}
\caption{(Color online) Supercurrent through the island as a function of voltage for various $\mathcal{K}_{coll}$ with $T=0.7\:T_C$. For the hysteretic curves the arrows indicate the direction the curve is traced when the bias voltage is varied.}
\label{fig:collsc}
\end{figure}

When the energy relaxation due to inelastic electron-electron scattering is taken into account, we are no longer able to obtain an explicit expression for $f_2$. In the left-right symmetric case we must instead solve the resulting integral equation
\begin{equation}
 f_2=\frac{g^{(-)}_{E+\mu}f_1+g^{(-)}_{E-\mu}f_3+\left(e^2\nu_2\Omega_2 R_1/g^{(-)}_{2,E}\right)J^{(ee)}_1[f_2]}{g^{(-)}_{E+\mu}+g^{(-)}_{E-\mu}}.
\end{equation}
The relaxation strength can be adjusted by varying the parameter $\mathcal{K}_{coll}=\kappa e^2\nu_2\Omega_2R_1$. The distribution function calculated for various values of $\mathcal{K}_{coll}$ is shown in Fig.~\ref{fig:colldists}. The energy distribution gradually relaxes towards a Fermi distribution upon increasing the strength of the relaxation. The influence of inelastic scattering to the supercurrent is shown in Fig.~\ref{fig:collsc} for a structure consisting entirely of one type of a superconductor. The enhancement of superconductivity is suppressed as the electron-electron collisions drive the electron temperature of the central island towards quasiequilibrium. With the strongest relaxation the cooling effect is completely lost and the supercurrent drops smoothly to zero as the voltage is increased. With the two largest strengths of relaxation the hysteresis is lost as well. At larger voltages the supercurrent is a non-monotonous function of $\mathcal{K}_{coll}$, as the supercurrent in quasiequilibrium ($\mathcal{K}_{coll}=\infty$) is significantly larger than with moderate relaxation. This can also be seen in the left part of Fig.~\ref{fig:colldists}, where the distribution function in quasiequilibrium is sharper compared to the distribution with $\mathcal{K}_{coll}=10$. This sharpness leads directly to a larger supercurrent. The curves with $\mathcal{K}_{coll}=1$ and $\mathcal{K}_{coll}=2$ show a small jump in the supercurrent at voltages over $eV=2$. This corresponds to a transition above which $\Delta_4>\Delta_2$. By choosing a setup where leads 4 and 5 have a different energy gap from the rest of the system, this peak could be seen at different voltages.

\subsection{Asymmetric structure}
Let us now examine an asymmetric situation, where $R_1\neq R_3$ or $\Delta_1\neq\Delta_3$. By solving the kinetic equations \eqref{eq:kineticeqL} and \eqref{eq:kineticeqT} without relaxation we obtain quite lengthy expressions for the odd and even parts of the distribution function
\begin{widetext}
\begin{align}
 Df_{L2}=&4e^2f^{(+)}_2\nu_2\Omega_2|\Delta_2|\left[G_1\left(\left(f_{L1}-f_{T1}\right)g^{(-)}_{1,E-\mu_1}+\left(f_{L1}+f_{T1}\right) g^{(-)}_{1,E+\mu_1}\right)+\right. \nonumber \\
 &\left.G_3\left(\left(f_{L3}-f_{T3}\right)g^{(-)}_{3,E-\mu_3}+\left(f_{L3}+f_{T3}\right) g^{(-)}_{3,E+\mu_3}\right)\right] \nonumber \\
 &-g^{(-)}_{2,E}\left[G_1\left(2f_{L1}G_1g^{(-)}_{1,E-\mu_1}+\left(f_{L1}+f_{L3}+f_{T1}-f_{T3}\right)G_3 g^{(-)}_{3,E-\mu_3}\right)g^{(-)}_{1,E+\mu_1} \right.\nonumber \\
 &\left.+G_3\left(2f_{L3}G_3g^{(-)}_{3,E-\mu_3}+\left(f_{L1}+f_{L3}-f_{T1}+f_{T3}\right)G_1g^{(-)}_{1,E-\mu_1}\right)g^{(-)}_{3,E+\mu_3}\right], \nonumber \\
 Df_{T2}=&-g^{(-)}_{2,E}\left[G_1\left(2f_{T1}G_1g^{(-)}_{1,E-\mu_1}+\left(f_{L1}-f_{L3}+f_{T1}+f_{T3}\right) G_3g^{(-)}_{3,E-\mu_3}\right)g^{(-)}_{1,E+\mu_1}\right. \nonumber \\
 &\left.+G_3\left(2f_{T3}G_3g^{(-)}_{3,E-\mu_3}+\left(-f_{L1}+f_{L3}+f_{T1}+f_{T3}\right)G_1g^{(-)}_{1,E-\mu_1}\right)g^{(-)}_{3,E+\mu_3}\right],
\end{align}
where
\begin{align}
 D=&4e^2f^{(+)}_2\nu_2\Omega_2|\Delta_2|\left[G_1\left(g^{(-)}_{1,E-\mu_1}+g^{(-)}_{1,E+\mu_1}\right)+G_3\left(g^{(-)}_{3,E-\mu_3}+g^{(-)}_{3,E+\mu_3}\right)\right] \nonumber \\
 &-2g^{(-)}_{2,E}\left(G_1g^{(-)}_{1,E-\mu_1}+G_3g^{(-)}_{3,E-\mu_3}\right)\left(G_1g^{(-)}_{1,E+\mu_1}+G_3g^{(-)}_{3,E+\mu_3}\right). \nonumber
\end{align}
\end{widetext}
The distribution functions depend on the volume, energy gap and normal state density of states at the Fermi level, but these can be included in dimensionless constants of the type $G/|\Delta|\nu\Omega e^2$. In the asymmetric case the potentials $\mu_1$ and $\mu_3$ must be chosen such that the electrical current is conserved. This implies the vanishing of the total net current into the island, i.e., $I^{1\to2}=I^{2\to3}$ calculated with Eq.~\eqref{eq:qpcurrent}.

\begin{figure}[tb]
\includegraphics[width=8.0cm]{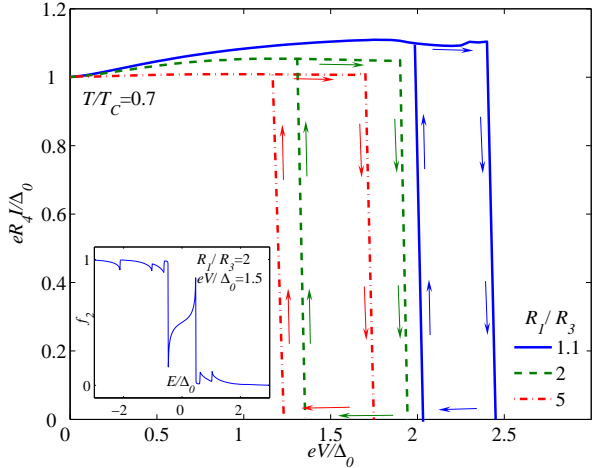}
\caption{(Color online) Supercurrent through the island as a function of voltage for different degrees of asymmetry in the SISIS control line. The arrows indicate the direction the curve is traced when the bias voltage is varied. The inset shows the distribution function on the island for a ratio $R_1/R_3=2$ at $eV/\Delta_0=1.5$. The distribution function exhibits small asymmetry due to finite $f_T$ as can be seen by the additional notch at negative energies.}
\label{fig:asymm}
\end{figure}
The supercurrent for $\chi_4-\chi_2=\pi/2$ and different ratios $R_1/R_3$ is shown in Fig.~\ref{fig:asymm}. In an asymmetric structure the magnitude of the order parameter seems to be close to its value in the symmetric case with a voltage of $eV=2\max\left(|\mu_1|,|\mu_3|\right)$. This is reasonable because the distribution function in the region $|E|>\max\left(\Delta_1+|\mu_1|,\Delta_3+|\mu_3|\right)$ is similar to the distribution in the symmetric structure as shown in the inset. Superconductivity is lost once $|\mu_1|>\Delta_2+\Delta_1$ or $|\mu_3|>\Delta_2+\Delta_3$. With high asymmetry ratios the potentials differ very much from $\pm eV/2$ and superconductivity is lost at a lower bias voltage compared to the symmetric structure. Also the hysteretic region is evident.
\begin{figure}[tb]
\includegraphics[width=8.0cm]{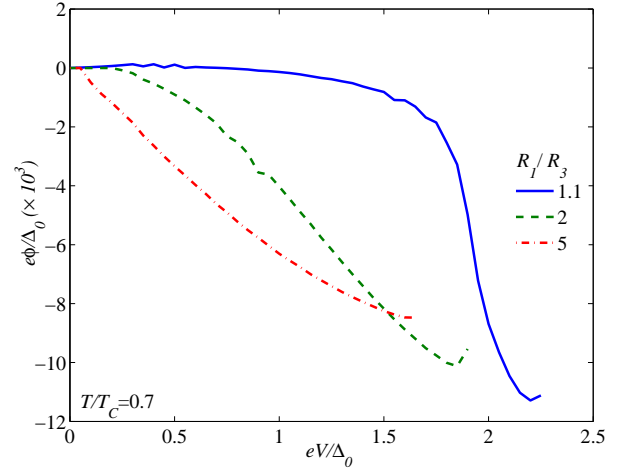}
\caption{(Color online) Potential difference between quasiparticles and the condensate for different degrees of asymmetry in the SISIS control line.}
\label{fig:elpots}
\end{figure}

Because the charge imbalance function, $f_T$, is finite, also the
latter part of Eq.~\eqref{eq:supercurrent} may contribute, depending
on the phase difference between the superconductors. Its magnitude
can be investigated by setting the phase difference to zero. In this
case the electric current is significantly smaller, of the order of
$10^{-3}\:eR_4I/\Delta_0$, and mostly due to quasiparticle current
induced by the charge imbalance. The charge imbalance leads to a
difference in the chemical potentials between quasiparticles and the
condensate. The potential difference is given by\cite{voutilainen}
\begin{equation}\label{eq:electricpotential}
 e\varphi=-\int_{-\infty}^\infty\frac{dE}{2}f_{T2}g^{(-)}_2.
\end{equation}
This quantity is shown in Fig.~\ref{fig:elpots} for the superconducting regime. The quasiparticle current depends linearily on this potential difference. The equation for the supercurrent also seems to imply a $\cos(\Delta\chi)$ dependence in the supercurrent. This deviation from the dc Josephson relation is negligible however, because the integral over the supercurrent term $f^{(+)}_2f^{(+)}_4$ is three orders of magnitude smaller than over the quasiparticle current term $g^{(-)}_2g^{(-)}_4$.

\section{Discussion}
According to our results there are several measurable features
present in a nonequilibrium, all-superconducting, tunnel structure.
The initial electron heating is seen as a strong suppression in
superconductivity of the central island when the tunnel structure is
biased with a low voltage. This is observable when the bath
temperature is slightly below the critical temperature of the
central island but well below the critical temperature of the
superconducting leads. The nonequilibrium cooling effect together
with the destruction of superconductivity at
$eV=2(\Delta_2+\Delta_L)$ should be observable with a wide range of
configurations. The accompanying hysteresis with low or nonexistent
relaxation can be seen in the supercurrent as well. The magnitude of
the energy gap could be directly measured with a quasiparticle
current probe, where the jump in the current happens at a probe
voltage of $eV=2(\Delta_2+\Delta_L)$.

Due to hysteresis the application of this structure as a transistor is unfeasible in states far from equilibrium. With moderate to strong relaxation the hysteresis is absent and does not hamper transistor-like operation. The sharp current-voltage characteristics giving rise to high differential current gain are lost with the strongest relaxation, however. If actual power gain were to be achieved, the Josephson junctions have to be operated in the dissipative regime and coupling to the environment should be taken into account in the calculations.

Small asymmetries of the order of ten percent in the system do not
give rise to qualitatively different behaviour. Asymmetries larger
than that begin to develop charge imbalance in the central island,
leading to different chemical potentials for the superconducting
condensate and quasiparticle excitations. This potential difference
can be observed in the quasiparticle current flowing to the island
from both reservoirs 4 and 5, when the phase difference accross the
Josephson junctions vanishes.

\begin{acknowledgments}
We thank J. Pekola for discussions. TTH is supported by the Academy of Finland and PV by the Finnish Cultural Foundation.
\end{acknowledgments}


\end{document}